\pdfoutput=1

\documentclass[preprint,5p,numbers,sort&compress]{elsarticle}

\usepackage{graphicx}
\usepackage{amssymb,amsmath}
\usepackage{natbib}
\usepackage{hyperref}
\hypersetup{pdftitle = The title of my PDF, pdfauthor = My name, pdfsubject= The subject, pdfkeywords = keyword1 keyword2 keyword3}
\hypersetup{colorlinks = true, linkcolor = blue, anchorcolor = red, citecolor = blue, filecolor = red, pagecolor = red, urlcolor = blue}

\journal{International Journal of Non-Linear Mechanics}

\begin{document}

\begin{frontmatter}
\title{Basins of convergence of equilibrium points in the generalized H\'{e}non-Heiles system}

\author[eez]{Euaggelos E. Zotos\corref{cor1}}
\ead{evzotos@physics.auth.gr}

\author[fld]{A. Ria\~{n}o-Doncel}

\author[fld]{F. L. Dubeibe}

\cortext[cor1]{Corresponding author}

\address[eez]{Department of Physics, School of Science,
Aristotle University of Thessaloniki,
GR-541 24, Thessaloniki, Greece}

\address[fld]{Facultad de Ciencias Humanas y de la Educaci\'{o}n,
Universidad de los Llanos, Villavicencio, Colombia}

\begin{abstract}
We numerically explore the Newton-Raphson basins of convergence, related to the libration points (which act as attractors of the convergence process), in the generalized H\'{e}non-Heiles system (GHH). The evolution of the position as well as of the linear stability of the equilibrium points is determined, as a function of the value of the perturbation parameter. The attracting regions, on the configuration $(x,y)$ plane, are revealed by using the multivariate version of the classical Newton-Raphson iterative algorithm. We perform a systematic investigation in an attempt to understand how the perturbation parameter affects the geometry as well as of the basin entropy of the attracting domains. The convergence regions are also related with the required number of iterations.
\end{abstract}


\end{frontmatter}

\section{Introduction}
\label{intro}

In every two-dimensional Hamiltonian system the determination of the fixed points is a necessary step in order to diagnose the global dynamical behavior of the system. The usual way to do so is by solving the algebraic system of equations
\begin{equation}
\frac{\partial V}{\partial x} = \frac{\partial V}{\partial y} = 0,
\label{sys}
\end{equation}
where $V$ denotes the effective potential function, associated to the physical system. In the majority of cases, it is
commonly necessary to solve numerically the system (\ref{sys}) via root finding algorithms, which use iterations and require one or more initial guesses of the root as starting values. In this realm, it is important to consider the possible appearance of two issues: the non-convergence of a given initial condition and the number of iterations needed to reach a given fixed point. These two issues can be treated by means of the so called basin of convergence, which gives us information about: (i) the total number of fixed points in the system, (ii) the non-converging initial conditions and (iii) the number of iterations needed to reach a given fixed point.

In a general sense, the basins of convergence can be understood as the set of initial guesses (conditions) that after iterations, via a root-finding algorithm, tend to a fixed point (see e.g., \cite{K04} for the case of complex polynomials) and must not be confused with the concept of basin of attraction, defined as the set of initial conditions that lead to
a specific attractor \cite{NY96}, because the last one is meaningless in the context of conservative Hamiltonian systems. As in the case of basins of attraction, the basins of convergence could be composed by more than one basin (associated to each fixed point) separated by boundaries which can be smooth or fractal curves. During the last years the basins of convergence have been investigated in many different dynamical systems, using the Newton-Raphson iterative method, e.g. for the Hill's problem with radiation and oblateness \cite{D10}, the Sitnikov problem \cite{DKMP12}, the restricted four-body problem \cite{BP11}, the Copenhagen problem with magnetized primaries \cite{KGK12}, the restricted four-body problem with oblate primaries \cite{KK14}, the restricted four-body problem with radiation pressure \cite{APHS16}, the planar circular restricted three-body problem with oblateness and radiation pressure \cite{Z16}, the pseudo-Newtonian planar circular restricted three-body problem \cite{Z17}, the ring problem of $N + 1$ bodies \cite{CK07,GKK09}, or even the restricted 2 + 2 body
problem \cite{CK13}.

The H\'{e}non-Heiles Hamiltonian is a two-dimensional time-independent dynamical system, originally proposed as a simplified version of the gravitational potential experimented by a star orbiting around an axially symmetric galaxy (see e.g., \cite{HH64}). An extension of this potential up to the fourth-order was performed by Verhulst \cite{V79}, while in \cite{DRDZ17} we generalized the H\'{e}non-Heiles potential up to the fifth-order. In the present paper, we examine the
basins of convergence in the fifth-order generalization of the H\'{e}non-Heiles Hamiltonian (in all that follows GHH). The GHH is highly nonlinear, so an adequate tool to analyze the convergence properties of iteration functions is the Newton-Raphson algorithm. On the other hand, the fractality of the basins will be analyzed through the basin entropy, a new measure introduced recently to quantify the uncertainty of basins (e.g. of escape, convergence or attraction) \cite{DWGGS16}.

The present paper has the following structure: the main properties of the dynamical system are presented in Section \ref{mod}. The parametric evolution of the position as well as the stability of the equilibrium points is investigated in Section \ref{param}. Section \ref{res} contains the most relevant results regarding the evolution of the Newton-Raphson
basins of convergence. In Section \ref{bas} we demonstrate how the basin entropy of the configuration $(x,y)$ convergence
planes evolves, as a function of the perturbation parameter. Our paper ends with Section \ref{conc}, where we summarize the main conclusions of this work.

\section{Properties of the dynamical system}
\label{mod}

Let us briefly recall the main properties of the generalized H\'{e}non-Heiles system (GHH). The corresponding potential is given by
\begin{align}
U(x,y) &= \frac{1}{2}\left(x^2 + y^2 \right) + x^2 y - \frac{y^3}{3} \nonumber\\
&+ \delta \left[x^4 y + x^2 y^3 - y^5 - \left(x^2 + y^2\right)^2 \right],
\label{pot}
\end{align}
where $\delta$ is a perturbation parameter. Note that when $\delta = 0$ the potential (\ref{pot}) is automatically reduced to that of the classical H\'{e}non-Heiles system.

Potential (\ref{pot}) is derived as a Taylor expansion up to the 5-th order of a general potential with axial and reflection symmetries. More information about the exact expansion and the derivation of the generalized potential is given in \cite{DRDZ17}.

The equations of motion read
\begin{equation}
\ddot{x} = - \frac{\partial U}{\partial x},  \ \ \
\ddot{y} = - \frac{\partial U}{\partial y},
\label{eqmot}
\end{equation}
where
\begin{align}
U_x(x,y) &= \frac{\partial U}{\partial x} = x \left(1 + 2 y\right) \nonumber\\
&+ 2 \delta x \left(2x^2 \left(y - 1\right) + y^2 \left(y - 2\right)\right), \nonumber\\
U_y(x,y) &= \frac{\partial U}{\partial x} = x^2 + y \left(1 - y\right) \nonumber\\
&+ \delta \left(x^4 + x^2 y \left(3y - 4\right) - y^2 \left(5y - 4\right)\right).
\label{der1}
\end{align}
In the same vein, the second order derivatives of the potential $U(x,y)$, which will be needed later for the computation of the multivariate Newton-Raphson iterative scheme are as follows
\begin{align}
U_{xx}(x,y) &= \frac{\partial^2 U}{\partial x^2} = 1 + 2y + 2\delta \left(6x^2 \left(y - 1\right) + y^2 \left(y - 2\right)\right), \nonumber\\
U_{xy}(x,y) &= \frac{\partial^2 U}{\partial x \partial y} = 2x + \delta \left(4x^3 + 2xy \left(3y - 4\right)\right), \nonumber\\
U_{yx}(x,y) &= \frac{\partial^2 U}{\partial y \partial x} = U_{xy}(x,y), \nonumber\\
U_{yy}(x,y) &= \frac{\partial^2 U}{\partial y^2} = 1 - 2y + 2\delta \left(x^2 \left(3y - 2\right) - 2y^2 \left(5y - 3\right)\right).
\label{der2}
\end{align}

The Hamiltonian, which dictates the motion of the test particle, is given by
\begin{equation}
H\left(x,y,\dot{x},\dot{y}\right) = U(x,y) + \frac{1}{2} \left(\dot{x}^2 + \dot{y}^2 \right) = E,
\label{ham}
\end{equation}
where $\dot{x}$ and $\dot{y}$ are the velocities, associated to the coordinates $x$ and $y$, respectively, while $E$ is the numerical value of the total orbital energy of the test particle, which is conserved.

\section{Parametric evolution of the equilibrium points}
\label{param}

The necessary and sufficient conditions that must be satisfied for the existence of coplanar equilibrium points, are
\begin{equation}
\dot{x} = \dot{y} = \ddot{x} = \ddot{y} = 0.
\label{cond}
\end{equation}
On the other hand, the positions of the libration points can be determined by numerically solving the system of equations
\begin{equation}
U_x(x,y) = U_y(x,y) = 0.
\label{sys2}
\end{equation}

\begin{figure*}[!t]
\centering
\resizebox{\hsize}{!}{\includegraphics{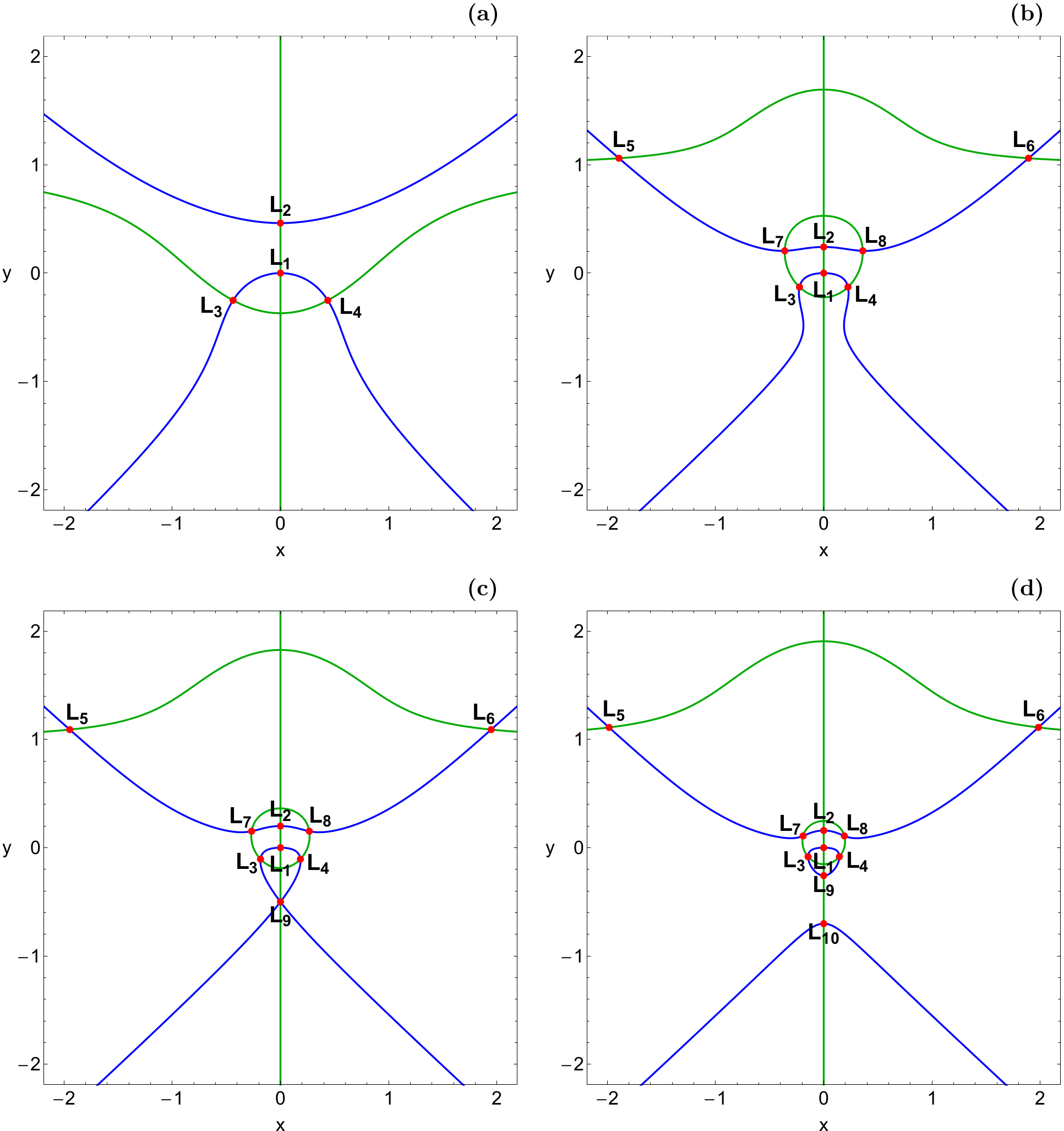}}
\caption{Positions (red dots) and numbering of the equilibrium points $(L_i$, $i = 1,..., 10)$ through the intersections of $U_x = 0$ (green) and $U_y = 0$ (blue), when (a-upper left): $\delta = 0.4$ (four equilibrium points), (b-upper right): $\delta = 2.5$ (eight equilibrium points), (c-lower left): $\delta = 4$ (nine equilibrium points) and (d-lower right): $\delta = 7$ (ten equilibrium points). (Color figure online).}
\label{lgs}
\end{figure*}

\begin{figure*}[!t]
\centering
\resizebox{0.8\hsize}{!}{\includegraphics{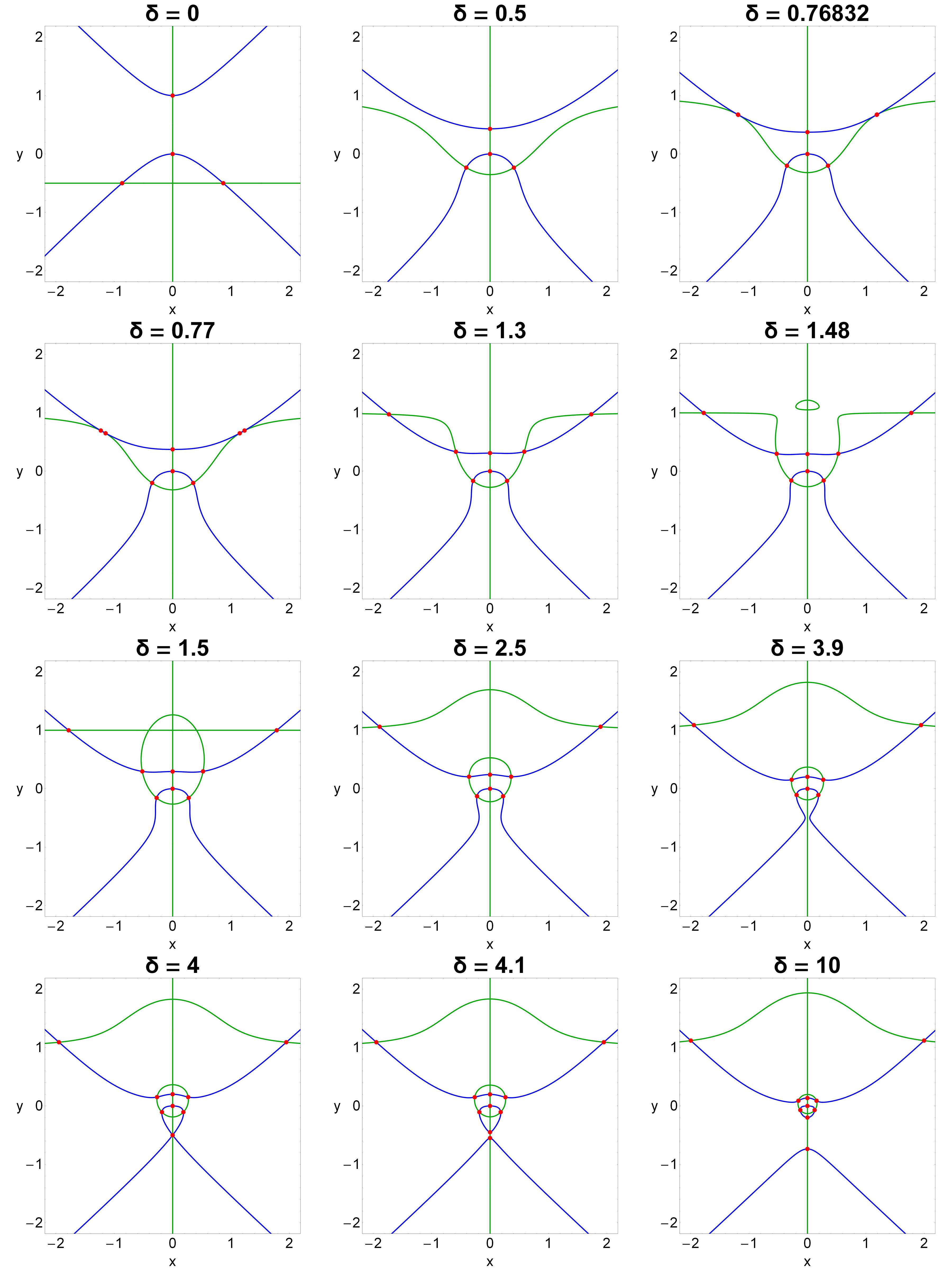}}
\caption{The variation of the positions of the equilibrium points (red dots) and the contours defined by the equations $U_x = 0$, $U_y = 0$, as a function of the perturbation parameter $\delta$. (Color figure online).}
\label{conts}
\end{figure*}

From Eqs. (\ref{pot}) and (\ref{sys2}), it can be easily noticed that the total number of libration points in the GHH is a function of the perturbation parameter $\delta$. In particular
\begin{itemize}
  \item When $\delta \in [0, 0.76831375]$ there exist four equilibrium points (see panel (a) of Fig. \ref{lgs}).
  \item When $\delta \in [0.76831376, 4)$ there exist eight equilibrium points (see panel (b) of Fig. \ref{lgs}).
  \item When $\delta = 4$ there are nine libration points (see panel (c) of Fig. \ref{lgs}).
  \item When $\delta > 4$ there exist ten equilibrium points (see panel (d) of Fig. \ref{lgs}).
\end{itemize}
Our numerical analysis suggests that the exact value of $\delta$ for which six equilibrium points are present is an irrational number and therefore it cannot be exactly determined.

As shown in panels (a)-(d) of Fig. \ref{lgs} for (a): $\delta = 0.4$, (b): $\delta = 2.5$, (c): $\delta = 4$, (d): $\delta = 7$, the intersection points of the curves corresponding to the first order derivatives (\ref{sys2}), shall denote the position of the libration points $L_i$, $i = 1,..., 10$. Furthermore, in Fig. \ref{conts} we present how the number and the exact positions of the equilibrium points evolve as a function of the value of the perturbation parameter.

\begin{figure}[!t]
\centering
\resizebox{\hsize}{!}{\includegraphics{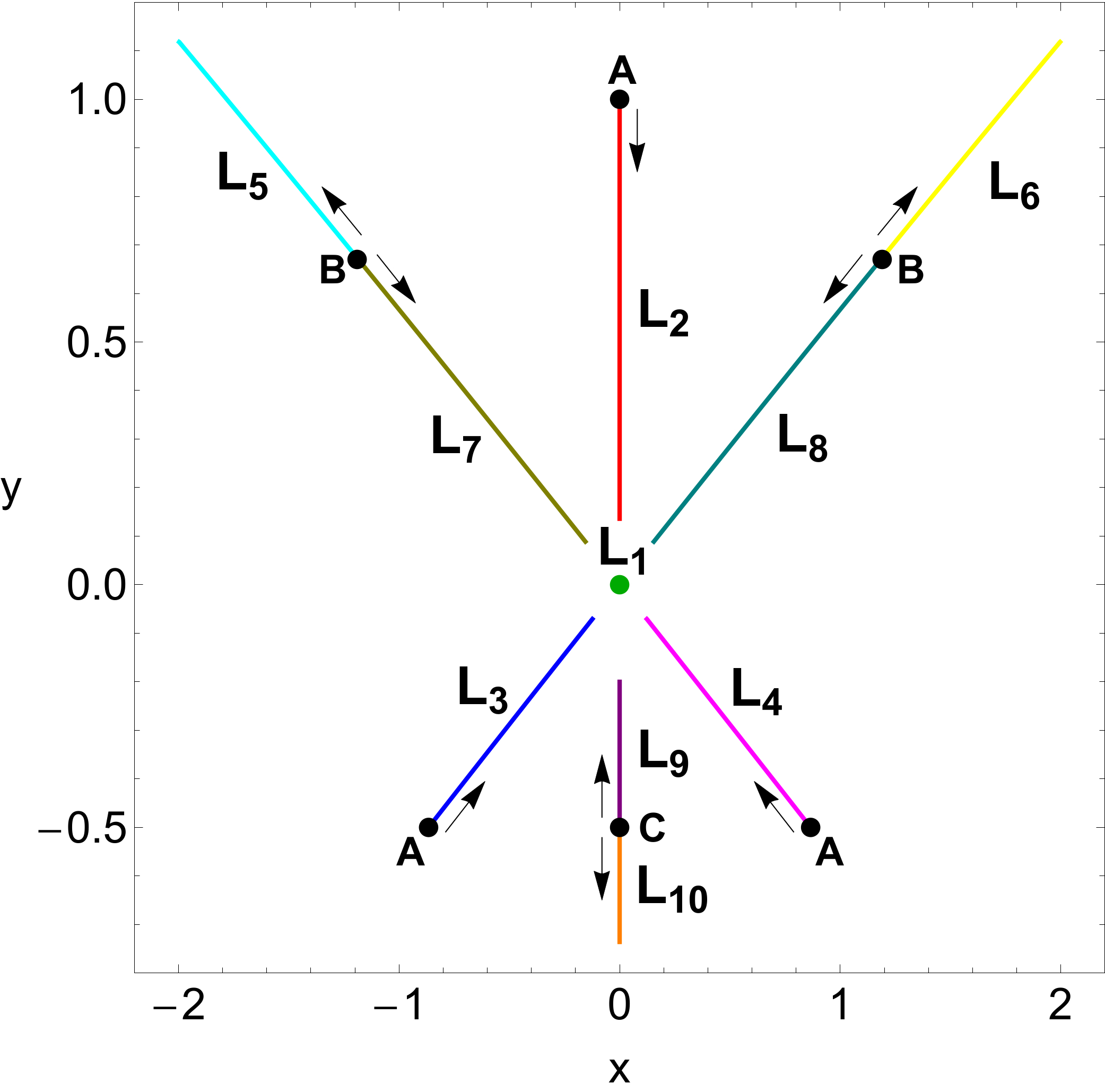}}
\caption{The evolution of the equilibrium points, $L_i$, $i = 1,..., 10$, in the GHH, when $\delta \in [0,10]$. The direction of displacement of the equilibrium points as the value of $\delta$ increases is indicated by the arrows. The black dots (points A, B, and C) correspond to $\delta \to 0$, $\delta = 0.76831376$, and $\delta = 4$, respectively. (Color figure online).}
\label{evol}
\end{figure}

In Fig. \ref{evol} we present the parametric evolution of the positions of the equilibrium points, on the configuration $(x,y)$ plane, when $\delta \in [0, 10]$. It is seen that the position of the central libration point $L_1$ remains unperturbed at the origin $(0,0)$, while the location of all the other equilibrium points changes linearly. As soon as $\delta \geq 0.76831376$ two pairs of libration points emerge, while one additional pair appears when $\delta > 4$. Our computations indicate that in the limit $\delta \to \infty L_2, L_3, L_4, L_7, L_8$, and $L_9$ tend to collide with $L_1$, while on the other hand $L_5$, $L_6$, and $L_{10}$ move far away from the center.

Following the procedure outlined in \cite{Z17}, once we know the positions of the libration points, we can determine their linear stability by means of the characteristic equation. To do so, we defined a dense, uniform sequence of $10^5$ values of $\delta$ in the interval $[0,10]$ and numerically solved the system (\ref{sys2}), thus determining the coordinates $(x_0,y_0)$ of the equilibrium points. The last step was to insert the coordinates of the equilibria into the characteristic equation and determine the nature of the four roots. The above-mentioned numerical analysis reveals that all the equilibrium points $L_i$, $i = 2,..., 10$ are always linearly unstable, while $L_1$ is always Lyapunov stable, when $\delta > 0$.

\section{The basins of convergence}
\label{res}

One of the most well-known numerical methods for finding successive approximations to the roots of nonlinear equations is the Newton-Raphson method. This method is applicable to systems of multivariate functions $f({\bf{x}}) = 0$ through the iterative scheme
\begin{equation}
{\bf{x}}_{n+1} = {\bf{x}}_{n} - J^{-1}f({\bf{x}}_{n}),
\label{sch}
\end{equation}
where $f({\bf{x_n}})$ denotes the system of equations, while $J^{-1}$ is the corresponding inverse Jacobian matrix. In our case Eqs. (\ref{sys2}) describe the system of the differential equations.

The above-mentioned iterative scheme can be decomposed for each coordinate $x$ and $y$, as follows
\begin{align}
x_{n+1} &= x_n - \left( \frac{U_x U_{yy} - U_y U_{xy}}{U_{yy} U_{xx} - U^2_{xy}} \right)_{(x_n,y_n)}, \nonumber\\
y_{n+1} &= y_n + \left( \frac{U_x U_{yx} - U_y U_{xx}}{U_{yy} U_{xx} - U^2_{xy}} \right)_{(x_n,y_n)},
\label{nrm}
\end{align}
where $x_n$, $y_n$ are the values of the $x$ and $y$ coordinates at the $n$-th step of the iterative process.

The philosophy behind the Newton-Raphson method is the following: An initial condition $(x_0,y_0)$, on the configuration plane activates the code, while the iterative procedure continues until an equilibrium point (attractor) is reached, with the desired predefined accuracy. If the particular initial condition leads to one of the libration points of the system it means that the numerical method converges for that particular initial condition. At this point, it should be emphasized that in general terms the method does not converge equally well for all the available initial conditions. The sets of the initial conditions which lead to the same root compose the so-called Newton-Raphson basins of convergence or even attracting domains/regions. Nevertheless, as pointed out in the Introduction section, it should be clarified that the Newton-Raphson basins of convergence should not be confused, by no means, with the basins of attractions which are present in systems with
dissipation.

From the iterative formulae of Eqs. (\ref{nrm}) it becomes evident that they should reflect some of the most intrinsic
properties of the Hamiltonian system. This is true if we take into account that they contain the derivatives of both first and second order of the effective potential $U(x,y)$.

A double scan of the configuration $(x,y)$ plane is performed for revealing the structures of the basins of convergence. In particular, a dense uniform grid of $1024 \times 1024$ $(x_0,y_0)$ nodes is defined which shall be used as initial conditions of the iterative scheme. The number $N$ of the iterations, required for obtaining the desired accuracy, is also monitored during the classification of the nodes. For our computations, the maximum allowed number of iterations is $N_{\rm max} = 500$, while the iterations stop only when an attractor is reached, with an accuracy of $10^{-15}$.

\begin{figure*}[!t]
\centering
\resizebox{\hsize}{!}{\includegraphics{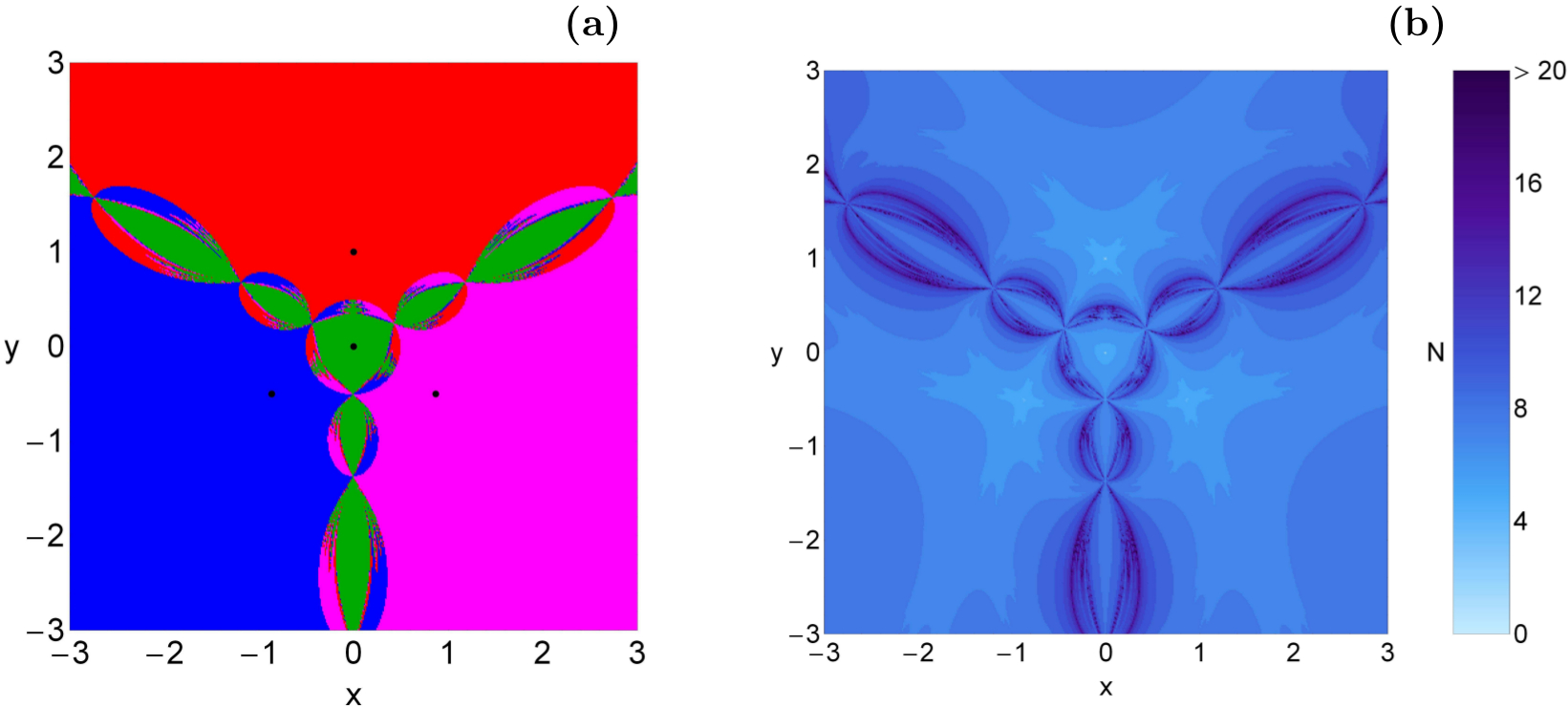}}
\caption{(a-left): The Newton-Raphson basins of convergence on the configuration $(x,y)$ plane for the classical H\'{e}non-Heiles system, where $\delta = 0$. The four equilibrium points are indicated by black dots. The initial conditions, leading to a certain equilibrium point, are marked using the following color code: $L_1$ (green); $L_2$ (red); $L_3$ (blue); $L_4$ (magenta); non-converging points (white). (b-right): The distribution of the corresponding number $N$ of required iterations for obtaining the Newton-Raphson basins of attraction shown in panel (a). (Color figure online).}
\label{sm}
\end{figure*}

The Newton-Raphson basins of convergence when $\delta = 0$ (which correspond to the classical H\'{e}non-Heiles system) are presented in panel (a) of Fig. \ref{sm}. Different colors are used for each basin of convergence, while the positions of
the four equilibrium points (attractors) are indicated by black dots. It is seen that the $2\pi/3$ symmetry of the system
is also displayed in the configuration $(x,y)$ plane, by the geometry of the attracting domains. The distribution of the corresponding number $N$ of iterations is given in panel (b) of the same figure, using tones of blue.

\begin{figure*}[!t]
\centering
\resizebox{0.7\hsize}{!}{\includegraphics{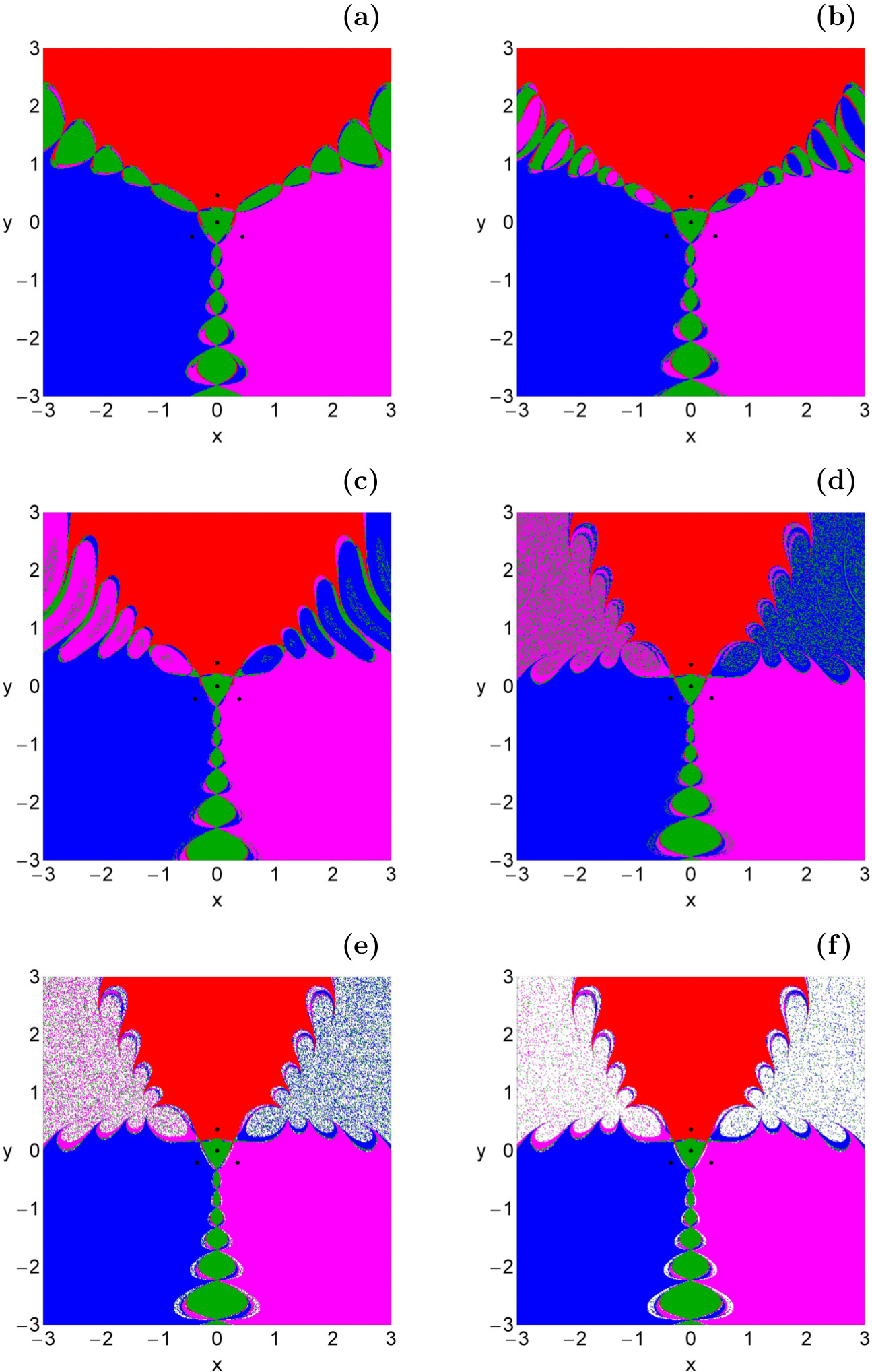}}
\caption{The Newton-Raphson basins of convergence on the configuration $(x,y)$ plane for the first case, where four equilibrium points are present. (a): $\delta = 0.40$; (b): $\delta = 0.45$; (c): $\delta = 0.60$; (d): $\delta = 0.75$; (e): $\delta = 0.7683$; (f): $\delta = 0.768313$. The positions of the equilibrium points are indicated by black dots. The color code, denoting the four attractors, is as follows: $L_1$ (green); $L_2$ (red); $L_3$ (blue); $L_4$ (magenta); non-converging points (white). (Color figure online).}
\label{c1}
\end{figure*}

\begin{figure*}[!t]
\centering
\resizebox{0.7\hsize}{!}{\includegraphics{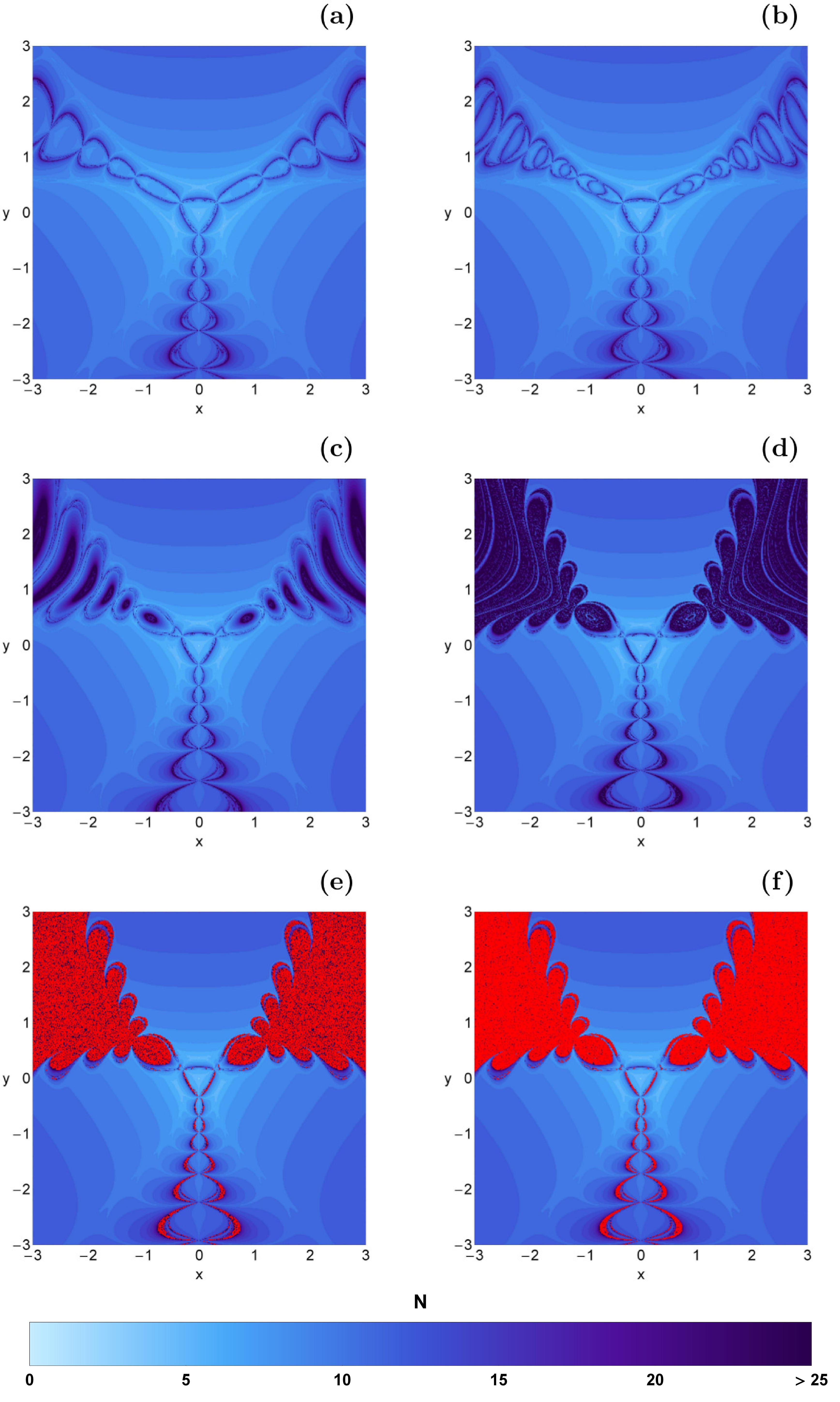}}
\caption{The distribution of the corresponding number $N$ of required iterations for obtaining the Newton-Raphson basins of convergence shown in Fig. \ref{c1}(a-f). The non-converging points are shown in red. (Color figure online).}
\label{n1}
\end{figure*}

\begin{figure*}[!t]
\centering
\resizebox{\hsize}{!}{\includegraphics{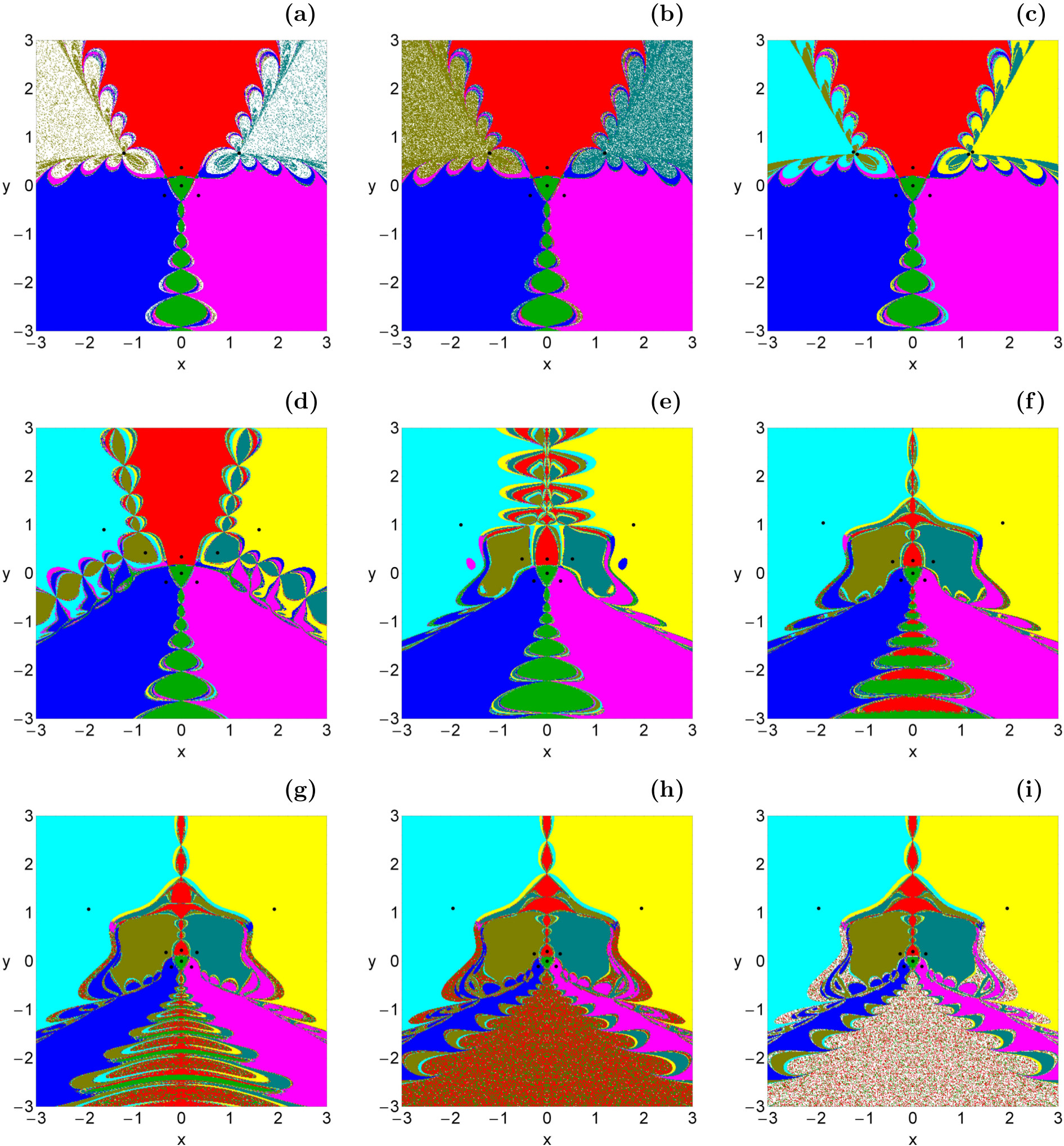}}
\caption{The Newton-Raphson basins of convergence on the configuration $(x,y)$ plane for the second case, where eight equilibrium points are present. (a): $\delta = 0.768314$; (b): $\delta = 0.76832$; (c): $\delta = 0.77$; (d): $\delta = 1$; (e): $\delta = 1.5$; (f): $\delta = 2$; (g): $\delta = 3$; (h): $\delta = 3.99$; (i) $\delta = 3.9999$. The positions of the equilibrium points are indicated by black dots. The color code, denoting the eight attractors, is as follows: $L_1$ (green);
$L_2$ (red); $L_3$ (blue); $L_4$ (magenta); $L_5$ (cyan); $L_6$ (yellow); $L_7$ (olive); $L_8$ (teal); non-converging points (white). (Color figure online).}
\label{c2}
\end{figure*}

\begin{figure*}[!t]
\centering
\resizebox{\hsize}{!}{\includegraphics{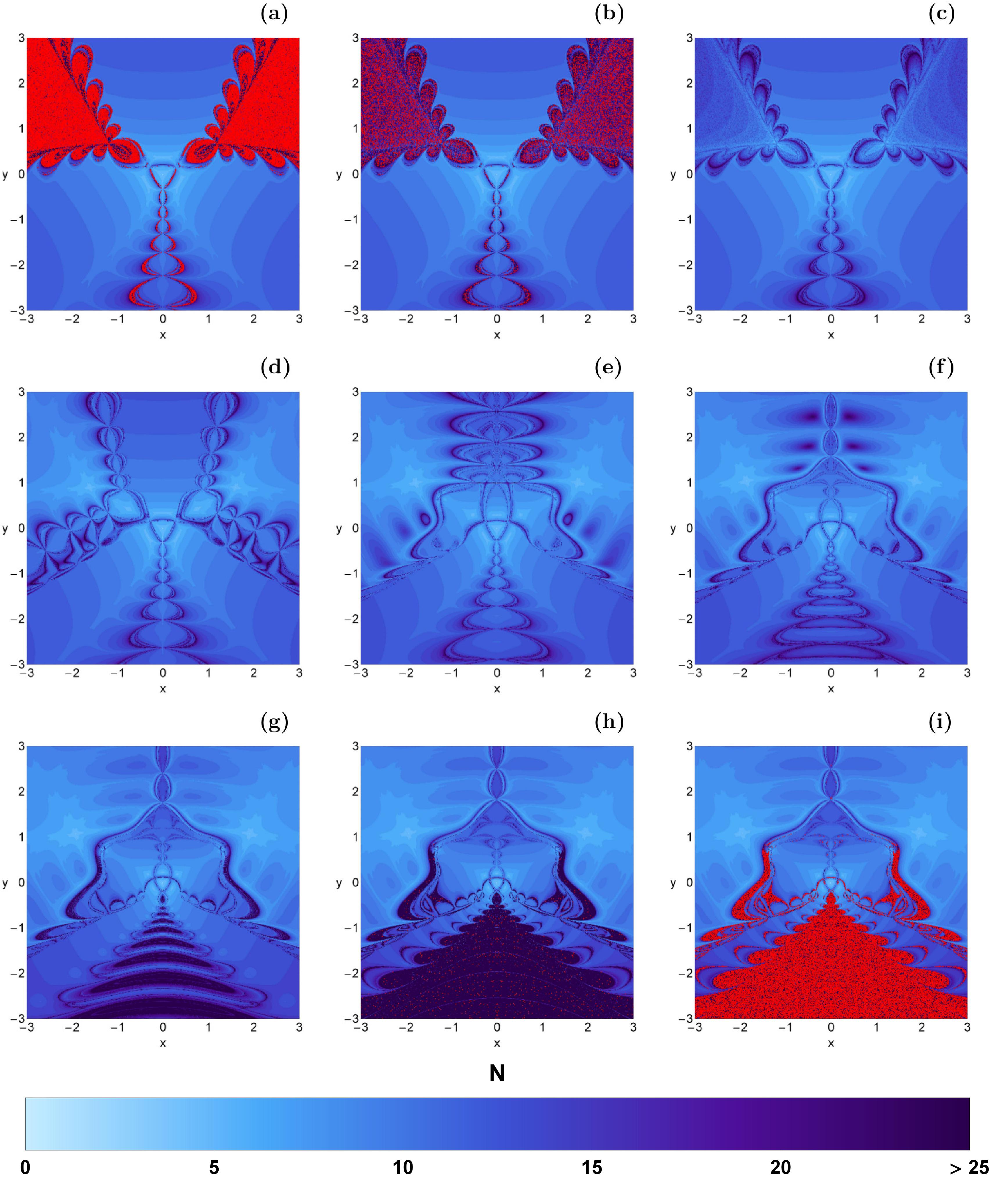}}
\caption{The distribution of the corresponding number $N$ of required iterations for obtaining the Newton-Raphson basins of convergence shown in Fig. \ref{c2}(a-f). The non-converging points are shown in red. (Color figure online).}
\label{n2}
\end{figure*}

\begin{figure*}[!t]
\centering
\resizebox{\hsize}{!}{\includegraphics{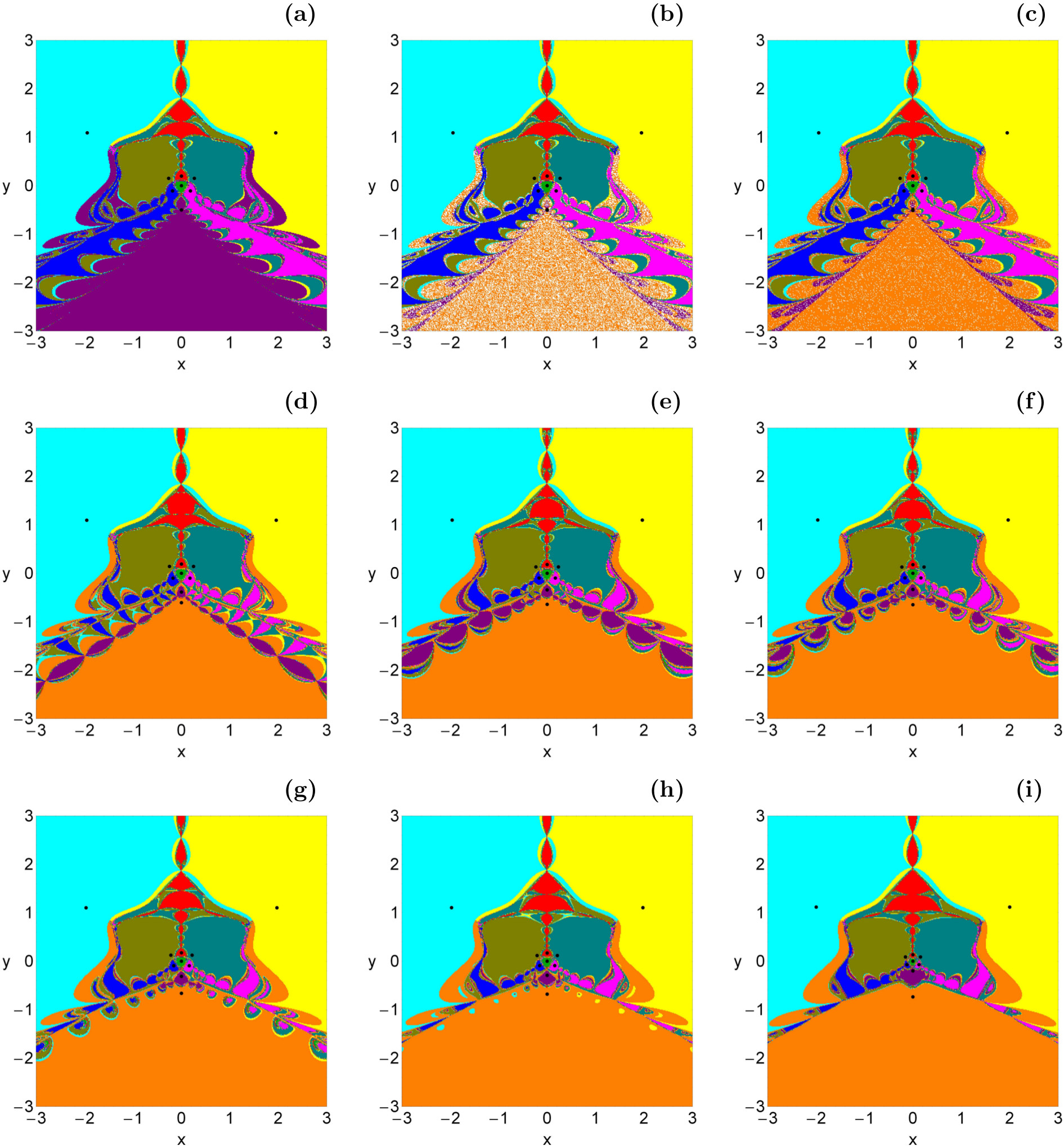}}
\caption{The Newton-Raphson basins of convergence on the configuration $(x,y)$ plane for the third case, where nine or ten equilibrium points are present. (a): $\delta = 4$; (b): $\delta = 4.001$; (c): $\delta = 4.005$; (d): $\delta = 4.5$; (e): $\delta = 5$; (f): $\delta = 5.15$; (g): $\delta = 5.5$; (h): $\delta = 6$; (i) $\delta = 10$. The positions of the equilibrium points are indicated by black dots. The color code, denoting the eight attractors, is as follows: $L_1$ (green);
$L_2$ (red); $L_3$ (blue); $L_4$ (magenta); $L_5$ (cyan); $L_6$ (yellow); $L_7$ (olive); $L_8$ (teal); $L_9$ (purple); $L_{10}$ (orange); non-converging points (white). (Color figure online).}
\label{c3}
\end{figure*}

\begin{figure*}[!t]
\centering
\resizebox{\hsize}{!}{\includegraphics{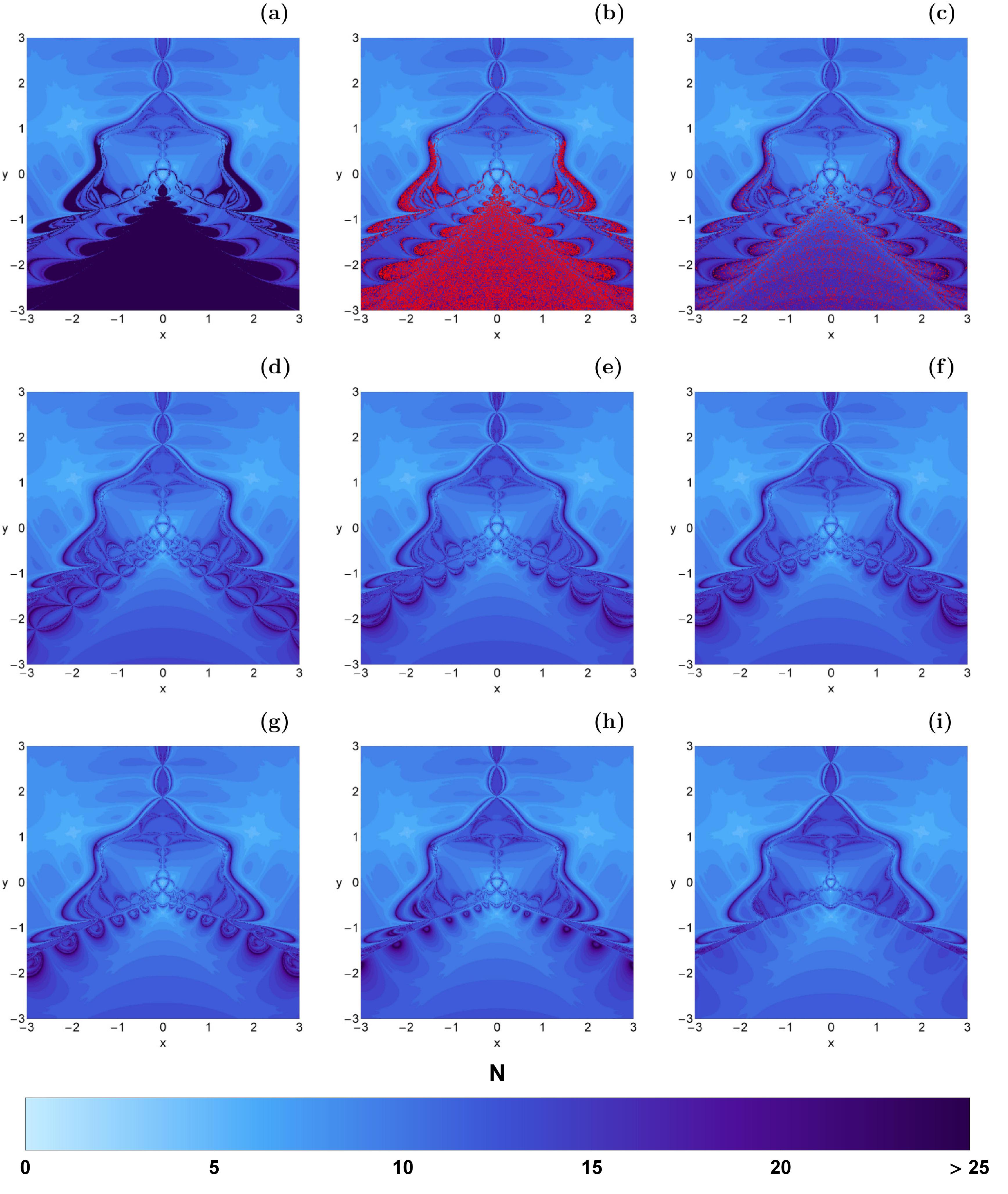}}
\caption{The distribution of the corresponding number $N$ of required iterations for obtaining the Newton-Raphson basins of convergence shown in Fig. \ref{c3}(a-f). The non-converging points are shown in red. (Color figure online).}
\label{n3}
\end{figure*}

In the following subsections, we will determine how the perturbation parameter $\delta$ affects the structure of the Newton-Raphson basins of convergence in the GHH, by considering two cases regarding the total number of the attractors
(equilibrium points). For the classification of the nodes in the configuration $(x,y)$ plane we will use color-coded diagrams (CCDs), in which each pixel is assigned a different color, according to the final state (attractor) of the corresponding initial condition.

\subsection{Case I: Four equilibrium points}
\label{ss1}

Our investigation begins with the first case where four equilibrium points are present, that is when $0 < \delta \leq 0.76831375$. In Fig. \ref{c1} we present the evolution of the basins of convergence for six values of the perturbation. We observe that as soon as $\delta > 0$ the $2\pi/3$ symmetry is destroyed. However, the attracting regions are still symmetric
with respect to the vertical $x = 0$ axis. As we proceed to higher values of the perturbation parameter the vast majority of the CCDs remains almost unperturbed. The only significant changes appear mainly in the vicinity of the basin boundaries. More precisely, the unpredictability in these regions increases rapidly, since they display a highly fractal geometry\footnote{With the term fractal we refer to the fractal-like geometry of the region, without conducting any additional calculations as in \cite{AVS01,AVS09}.}. When $\delta = 0.7683$ (see panel (e) of Fig. \ref{c1}) and also when $\delta = 0.768313$ (see panel (f) of Fig. \ref{c1}) we detected a considerable amount of non-converging initial conditions. Additional numerical calculations reveal that these initial conditions eventually do converge to one of the available attractors, but only after an extremely high number of iterations $N \gg 500$.

The distribution of the corresponding number $N$ of iterations is provided, using tones of blue, in Fig. \ref{n1}(a-f). We see that initial conditions inside the basins of attraction converge relatively quickly $(N < 15)$, while the long-lasting
converging points $(N > 20)$ are those in the vicinity of the basin boundaries.

\subsection{Case II: Eight equilibrium points}
\label{ss2}

In the case where $0.76831375 \leq \delta < 4$, there are eight equilibrium points on the configuration $(x,y)$ plane. In Fig. \ref{c2}(a-i) we present the Newton-Raphson basins of convergence for nine values of the perturbation parameter. In panels (a) and (b) of Fig. \ref{c2}, which correspond to $\delta = 0.768314$ and $\delta = 0.76832$, respectively, we observe, once more, the presence of non-converging initial conditions. However, in this case our analysis strongly suggests
that these initial conditions should be true non-converging points, taking into account that they do not show any numerical
sign of convergence, not even after 50000 iterations. For the same two values of the perturbation parameter we also observed another interesting phenomenon. In particular, for $\delta = 0.768314$ and $\delta = 0.76832$ we did not find any initial conditions that correspond to equilibrium points $L_5$ and $L_6$. We believe that the complete absence of converging points to attractors $L_5$ and $L_6$ is somehow related with the presence of non-converging points. For $\delta = 3.99$ (see panel (h) of Fig. \ref{c2}) and $\delta = 3.9999$ (see panel (i) of Fig. \ref{c2}) we also detected non-converging initial conditions. However, when we increased the allowed number of iterations to $N_{\rm max} = 50000$ all these initial conditions converged, sooner or later, to one of the first two attractors of the system ($L_1$ or $L_2$).

The distribution of the corresponding number $N$ of iterations needed to achieve the desired accuracy in our computations
is presented in Fig. \ref{n2}(a-i). Looking at panels (g) and (h) of Fig. \ref{n2} it is evident that the required number of
iterations, for initial conditions inside the basins located at the lower part of the CCDs, increases rapidly as we approach the second critical value of the perturbation parameter.

\subsection{Case III: Nine or ten equilibrium points}
\label{ss3}

The last case under consideration corresponds to the region $\delta \geq 4$, where there are nine or ten equilibrium points. In Fig. \ref{c3} we present, through the corresponding CCDs, the Newton-Raphson basins of convergence for nine values of the perturbation parameter. When $\delta = 4.001$, that is a value of the perturbation parameter just above the second critical value, we see that the lower part of the CCD (see panel (b) of Fig. \ref{c3}) is occupied by a highly fractal mixture of non-converging points and initial conditions that lead to the attractor $L_{10}$. However, as the value of $\delta$ increases, thus moving away from the critical level, the amount of non-converging points decreases rapidly and by $\delta = 0.77$ (see panel (d) of Fig. \ref{c3}) they have completely disappeared. Our numerical experiments indicate that these initial conditions must be true non-converging points, since their portion, in each CCD, remains unperturbed by the shift of the number of allowed iterations up to $_{\rm Nmax} = 50000$. In general terms it is seen that the configuration $(x,y)$ plane is dominated by well-formed unified basins of convergence (corresponding mainly to libration points $L_5$, $L_6$, and $L_{10}$). Moreover, the most noticeable change with increasing value of $\delta$ is the fact that basin boundaries become less noisy (fractal).

The corresponding distributions of the required number $N$ of iterations are given in Fig. \ref{n3}(a-i). It is interesting to note in panel (a) of Fig. \ref{n3} that the required iterations number, corresponding to the equilibrium point (attractor) $L_9$, is high $N > 25$. We believe that this behavior can be explained, in a way, if we take into account that $\delta = 4$ is a critical value and, as we have seen so far, near the critical values of the perturbation parameter the iterations number grow significantly.

\section{Parametric evolution of the basin entropy}
\label{bas}

As noted in the introduction, the basin entropy concept was recently introduced in \cite{DWGGS16} as a new quantitative measure of the uncertainty of a given basin (e.g. of escape, convergence or attraction). The idea behind the method is to subdivide the phase space into $N$ small cells, each of which containing at least one of the total number of final states $N_A$. Since the probability to find a state $j$ in the $i-$th cell is defined as $p_{i,j}$, the entropy for the $i-$th cell can be calculated by means of the Gibbs entropy as follows
\begin{equation}
S_{i} = \sum_{j=1}^{N_{A}}p_{i,j}\log\left(\frac{1}{p_{i,j}}\right).
\label{si}
\end{equation}
Then, the basin entropy is calculated as the average entropy for the total number of cells $N$, i.e.,
\begin{equation}
S_{b} = \frac{1}{N}\sum_{i=1}^{N} S_{i}=\frac{1}{N}\sum_{i=1}^{N} \sum_{j=1}^{N_{A}}p_{i,j}\log\left(\frac{1}{p_{i,j}}\right).
\label{sb}
\end{equation}
Finally, it is worth mentioning that there is a strong dependence between the total number of cells $N$ and the result for basin entropy, such that for larger values of $N$ a more precise value of $S_b$ is obtained. This technical problem can be tackled by randomly selecting small cells in phase space through a Monte Carlo procedure (see e.g., \cite{DWGGS16,DGGWS17}). Following this approach, we find that in our particular problem, the final value for the basin entropy does not change for $N > 2\times10^{5}$ cells, so in all cases, we have used $N=2.5 \times10^{5}$ cells.

\begin{figure}[!t]
\centering
\resizebox{\hsize}{!}{\includegraphics{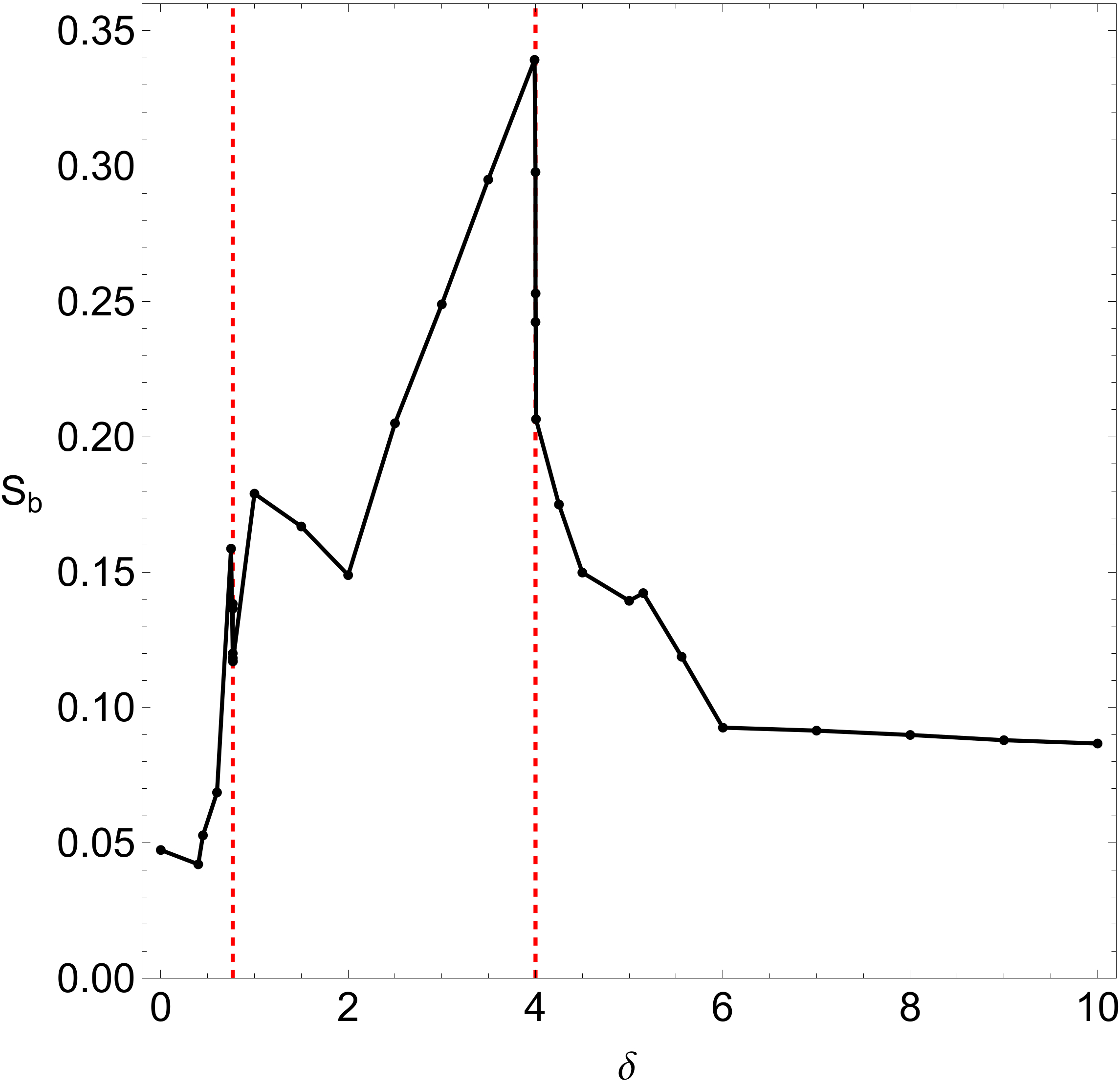}}
\caption{Evolution of the basin entropy $S_b$, as a function of the perturbation parameter $\delta$. The vertical dashed red lines indicate the critical values of $\delta$, where the total number of equilibrium points changes.}
\label{be}
\end{figure}

In Fig. \ref{be}, we present the parametric evolution of the basin entropy for different values of the perturbation parameter $\delta$, with $\delta \in [0, 10]$. Our results suggest that the basin entropy increases in the interval $\delta \in [0, 4]$, while for larger values of $\delta$ the basin entropy decreases almost monotonically. Particular attention deserve the critical values $\delta \approx 0.7683$ and  $\delta=4$, where small variations in the perturbation parameter give place to huge changes in the basin entropy. This result can be explained by considering that close to $\delta \approx 0.7683$ the system increases the number of roots from 4 to 8, i.e., the number of final states $N_{A}$ is modified and hence the value of Eq. (\ref{sb}) is also significantly modified. Similarly, close to $\delta = 4$ the number of roots change from 8 to 10 modifying the value of the basin entropy. On the other hand, it is important to note that the smaller and larger values of the basin entropy correspond to $\delta$ approximately  0 and 4, respectively. It means that the unpredictability associated to the NR basins of convergence for the classical H\'enon-Heiles system is smaller in comparison with the Generalized H\'enon-Heiles system.\footnote{Note that the differences for $\delta = 0$ and $\delta \approx 0.5$ are about 0.01, while the difference between $\delta = 0$ and $\delta \approx 4$ is approximately 0.3, i.e. 30 times greater.}

\section{Discussion and conclusions}
\label{conc}

We numerically explored the basins of convergence, related to the equilibrium points, in the generalized H\'{e}non- Heiles system. More precisely, we demonstrated how the perturbation parameter $\delta$ influences the position as well as the linear stability of the libration points. The multivariate version of the Newton-Raphson iterative scheme was used for revealing the corresponding basins of convergence on the configuration $(x,y)$ plane. These attracting domains play a significant role, since they explain how each point of the configuration plane is attracted by the libration points of the system, which act, in a way, as attractors. We managed to monitor how the Newton-Raphson basins of convergence evolve as a function of the perturbation parameter. Another important aspect of this work was the relation between the basins of convergence and the
corresponding number of required iterations.

To our knowledge, this is the first time that the Newton-Raphson basins of convergence, in the generalized H\'{e}non-Heiles system, are numerically investigated in a systematic manner. On this basis, the presented results are novel and this is exactly the contribution of our work.

The following list contains the most important conclusions of our numerical analysis.
\begin{enumerate}
  \item The stability analysis suggests that most of the equilibrium points of the system are always linearly unstable, when $\delta \geq 0$. Only the central libration point $L_1$ is Lyapunov stable, for the same values of the perturbation parameter.
  \item The attracting domains, associated to the libration points, extend to infinity, in all studied cases. Always the convergence diagrams, on the configuration $(x,y)$ plane are symmetrical, with respect to the vertical $x = 0$ axis. In the case of the classical H\'{e}non-Heiles system $(\delta = 0)$ there is an additional $2\pi/3$ symmetry.
  \item Near the critical values of the perturbation parameter, we detected the existence of non-converging initial conditions. Further numerical calculations revealed that in cases just below the critical values of $\delta$ these initial conditions do converge to one of the attractors but only after a considerable amount of iterations $(N \gg 500)$. On the other hand, in all cases just above the critical values of $\delta$ the corresponding initial conditions must be true non-converging points, since they do not display any numerical sign of convergence, not even after 50000 iterations.
  \item As expected, the multivariate Newton-Raphson method was found to converge very fast $(0 \leq N < 10)$ for initial conditions close to the equilibrium point and very slow $(N \geq 25)$ for initial conditions of dispersed points lying either in the vicinity of the basin boundaries, or between the dense regions of the libration points.
  \item The lowest value of the basin entropy was found near $\delta = 0$, while on the other hand the highest value of $S_b$ was measured near $\delta = 4$. This implies that the unpredictability, regarding the attracting regions, in the classical H\'{e}non-Heiles system is considerably smaller with respect to the generalized H\'{e}non-Heiles system.
\end{enumerate}

A double precision numerical code, written in standard \verb!FORTRAN 77! \citep{PTVF92}, was used for the classification of the initial conditions into the different basins of convergence. In addition, for all the graphical illustration of the paper we used the latest version 11.2 of Mathematica$^{\circledR}$ \citep{W03}. Using an Intel$^{\circledR}$ Quad-Core\textsuperscript{TM} i7 2.4 GHz PC the required CPU time, for the classification of each set of initial conditions, was about 5 minutes.

In the future, it would be very interesting to use other types of iterative schemes and compare the similarities as well as the differences on the corresponding basins of attraction. In particular, using iterative methods of higher order, with respect to the classical Newton-Raphson method, would be an ideal starting point. This would certainly lead to useful, and perhaps unexpected, results in the very active field of attracting domains of equilibrium points in dynamical systems.

\section*{Acknowledgments}
\footnotesize

FLD and ARD acknowledge financial support from Universidad de los Llanos, under Grant No. CDP 2478. FLD gratefully acknowledges the financial support provided by COLCIENCIAS, Colombia, under Grant No. 8840. The authors would like to express their warmest thanks to the two anonymous referees for the careful reading of the manuscript and for all the apt suggestions and comments which allowed us to improve both the quality and the clarity of the paper.

\end{document}